\def\umass{1}
\def\icts{2}
\def\upc{4}
\def\ieec{5}
\def\exeter{3}
\def\itc{6}

\documentclass[twocolumn, trackchanges]{aastex6}
\usepackage{natbib}

\usepackage{graphicx, verbatim}
\usepackage[space]{grffile}
\usepackage{latexsym}
\usepackage{amsfonts,amsmath,amssymb}
\usepackage{url}
\usepackage[utf8]{inputenc}
\usepackage{fancyref}

\usepackage{ulem} 
\usepackage{color} 

\begin{document}

\title{Double-Degenerate Carbon-Oxygen and Oxygen-Neon White Dwarf Mergers: A New Mechanism for Faint and Rapid Type Ia Supernovae}



\author{
Rahul Kashyap,\altaffilmark{\umass, \icts}
Tazkera Haque,\altaffilmark{\umass}
Pablo~Lor\'en-Aguilar,\altaffilmark{\exeter}
Enrique~Garc\'ia-Berro,\altaffilmark{\upc,\ieec}
Robert Fisher\altaffilmark{\umass, \itc}
}


\altaffiltext {\umass} {Department of Physics, University of Massachusetts Dartmouth, 285 Old Westport Road, North Dartmouth, Ma. 02740, USA}
\altaffiltext{\icts} {International Centre for Theoretical Sciences-TIFR, Survey no. 151, Shivakote,  Bengaluru - 560089, Karnataka, India}

\altaffiltext{\exeter}{School of Physics, University of Exeter, Stocker Road, Exeter EX4 4QL.}

\altaffiltext{\upc}{Departament de F\'{i}sica Aplicada, Universitat Polit\`{e}cnica de Catalunya, c/Esteve Terrades, 5, 08860 Castelldefels, Spain.}

\altaffiltext{\ieec}{Institut d'Estudis Espacials de Catalunya, Ed. Nexus-201, c/Gran Capit\`a 2-4, 08034 Barcelona, Spain.}

\altaffiltext {\itc} {Institute for Theory and Computation, Harvard-Smithsonian Center for Astrophysics, 60 Garden Street, Cambridge, MA 02138, USA}

\begin{abstract}

Type Ia supernovae (SNe Ia)  originate from the thermonuclear explosion of carbon-oxygen white dwarfs (CO WDs), giving rise to  luminous optical transients.  A relatively common variety of subluminous SNe Ia events, referred to as SNe Iax, are believed to arise from the failed detonation of a CO WD.  In this paper, we explore failed detonation SNe Ia in the context of the double-degenerate channel of merging white dwarfs. In particular, we have carried out the first fully three-dimensional simulation of the merger of a ONe WD with a CO WD. While the hot, tidally-disrupted carbon-rich disk material originating from the CO WD secondary is readily susceptible to detonation in such a merger, the ONe WD primary core is not. This merger yields a failed detonation, resulting in the ejection of a small amount of mass, and leaving behind a kicked, super-Chandrasekhar ONe WD remnant enriched by the fallback of the products of nuclear burning. The resulting outburst is a rapidly-fading optical transient with a small amount of radioactive $^{56}$Ni powering the light curve. Consequently, the ONe-CO WD  merger naturally produces a very faint and rapidly-fading transient, fainter even than the faintest Type Iax events observed to date, such as SN $2008$ha and SN 2010ae. More massive ONe primaries than considered here may produce brighter and longer-duration transients.

\end{abstract}


\section{Introduction}
\label {introduction}

Thermonuclear, or Type Ia supernovae (SNe Ia), are luminous optical transients characterized by the absence of hydrogen and helium lines, and the presence of a strong silicon P Cygni feature near maximum light \citep{Harkness1990}. 
The absence of  hydrogen down to very stringent mass limits provides a strong clue that the stellar progenitor undergoing the thermonuclear explosion is a white dwarf (WD) star.
A single CO WD with low rotation is typically born with a mass around 0.6 $M_{\odot}$ \citep{1998A&A...338..563H} and is not susceptible to  nuclear ignition in isolation. Consequently, SNe Ia are almost universally believed to arise in binary, or possibly triple, systems.  
One model invoking a degenerate sub-Chandrasekhar secondary WD is referred to as the double-degenerate (DD) channel. The merger of these two WDs may lead to conditions favorable to the detonation of carbon and hence a SNe Ia \citep{webbink84,Iben_Tutukov_1984}. 


Recent observations of a distinct class of astronomical subluminous transients, known as SNe Iax \citep{2013ApJ...767...57F}, pose challenges to current models. This class currently has more than fifty confirmed members, with a rate of $5$ to $30$ percent of the overall rate of SNe Ia \citep{2013ApJ...767...57F,2011MNRAS.412.1473L,2015ApJ...799...52W}. Although spectroscopically similar to SNe Ia, SNe Iax show distinctively lower maximum light velocities ($\sim 2000$ to $8000$ km/s) compared to those of SNe Ia ($\sim 10000$ to $15000$ km/s) \citep{2013ApJ...767...57F, 2015MNRAS.450.3045K, 2015A&A...573A...2S}, lower peak absolute magnitudes, lower luminosities with fast-declining light curves, a larger dispersion in the luminosity-width relationship \citep{2013ApJ...767...57F, 2014A&A...561A.146S,2010ApJ...720..704M}, and the  presence of narrow permitted Fe, Ca, and Na lines with P-Cygni profiles in the late-time spectra \citep{2013ApJ...767...57F, 2006AJ....131..527J, 2008ApJ...680..580S}. SNe Iax exhibit a range of peak magnitudes extending from $M_{\rm max}^{V} \approx$  $-18.1$ mag, observed for the brightest SNe Iax member SN 2005hk \citep{2006PASP..118..722C, 2007PASP..119..360P, 2008ApJ...680..580S} to the faintest event to date, SN 2008ha, with a peak magnitude of $M_{\rm max}^{V} \approx$ $-14.2$ mag \citep{foleyetal09b}. The progenitor channels and explosion mechanisms proposed so far to explain SNe Iax events are briefly outlined in \citet{2015MNRAS.450.3045K}. To date, the most favored models that best fit the observed properties of SNe Iax  are  deflagrations in a near-$M_{\rm ch}$ helium-accreting WD, failing to completely disrupt the WD \citep{jordanetal12b, kromeretal13,foleyetal09b}.

With recent improved stellar evolutionary models for super-AGB stars, a new variant of the single-degenerate channel, invoking a  ONe or hybrid CONe WD primary WD, has been proposed as a possible mechanism for SNe Iax events \citep{2015MNRAS.450.3045K}. In this proposed mechanism, the ONe WD accretes to near the Chandrasekhar mass limit, with off-center deflagration.  \citet {canalsetal18} have also recently carried out population synthesis models of the merger of ONe and CO WDs during common envelope evolution and found that their rates are less than normal SNe Ia, but could account at least partially for peculiar SNe Ia. In this paper, we consider a variant of the double-degenerate channel, in which a ONe WD merges with a massive CO WD. 


\subsection{ONe and Hybrid CONe WDs as SNe Ia Progenitors}
WDs are the final evolutionary state of nearly 97\% of the stars in a Milky-way like galaxy \citep{1538-3873-113-782-409}.
Different composition WDs can be produced from intermediate-mass stellar progenitors in the mass range $\approx$ $6.5-12$ $M_{\odot}$, which evolve through the super asymptotic giant branch (super-AGB) phase near the end of their lives.
If a super-AGB red giant has insufficient mass to raise the core temperature to the level required to ignite carbon, it will expel its outer envelope, and leave behind a CO core WD. Stellar evolutionary calculations suggest the CO core of an AGB star can possess a maximum mass of $\approx$1.1$M_{\odot}$, depending somewhat upon parameters including the stellar progenitor metallicity and the carbon-burning rate \citep{2014MNRAS.440.1274C}.

The carbon burning phase starts in an AGB star if the CO core mass of the star exceeds $\sim$1.1$M_{\odot}$ \citep{arnett69}.    Super-AGB stars in a critical mass range sufficient to ignite core C, but beneath the oxygen-burning limit \citep {schwabetal16}, will give rise to an oxygen-neon (ONe) or a hybrid carbon-oxygen-neon (CONe) core WD remnant \citet{refId0}, \citet{Doherty2017},  and \citet{ 2015ApJ...807..184F}.  In particular, an ONe core is produced when the carbon-burning flame proceeds all the way to the center of the star, and a hybrid CONe is produced when the carbon-burning flame does not make it completely to the center. 

 


The explosion mechanism of such hybrid CONe core WDs has been a subject of interest to researchers because of their wide range of potential fates as SNe explosions \citep {denissenkovetal14}. In a close binary system, these stars can more readily approach $M_{\rm ch}$ by accretion than lower-mass CO WDs, possibly giving rise to a thermonuclear explosion.  In particular, because of higher ignition temperatures of heavier O and Ne nuclei, in comparison to C, these systems may produce a failed complete detonation. \citet{2016A&A...589A..38B}, and \citet{2016ApJ...832...13W} explored the possible explosion mechanism of these hybrid core WDs, varying the initial size of core, and the explosion mechanism -- either a pure deflagration or a deflagration to detonation transition -- to demonstrate a characteristic low signature of $^{56}$Ni and lower kinetic energy in the ejecta, compared to pure CO core WDs. \citet{2015MNRAS.450.3045K} proposed a new mechanism involving a pure deflagration in the near-$M_{\rm ch}$ hybrid CONe WD  to explain the observational signatures of the faintest SNe Iax event SN 2008ha. Hence the failed detonation of CONe core hybrid WDs provides a possible explanation for subluminous SNe Iax events. 

\citet {jonesetal16} explored the fate of near-$M_{\rm ch}$ ONe WDs in the single-degenerate channel, finding that a strong deflagration could eject up to a solar mass of material, likely producing a subluminous SN Ia, and a bound ONeFe WD remnant. 
In the context of the double-degenerate channel, \citet {danetal15} explored a wide range of double-degenerate mergers as possible progenitors of SNe Ia, including one model consisting of a 0.95 $M_{\odot}$ CO WD with a 1.15 $M_{\odot}$ ONe WD. In particular, \citet {danetal15} found that this system failed to detonate either the envelope or the core as a natural consequence of their numerical simulations, initially evolved in 3D SPH and then remapped into an axisymmetric 2D {\tt FLASH} simulation. Only upon the artificial ignition of a hot spot in this model did they find a detonation of the ONe core and the production of 0.82 $M_{\odot}$ of $^{56}$Ni. 

In this paper, we report on results from a high-resolution, fully three-dimensional numerical simulation of the merger of a ONe WD with a CO WD through the double degenerate channel.  Analytic estimates for detonation initiation for this merger model are outlined in the following section.

\subsection {Thermonuclear Transients During Mergers of ONe and Massive CO WDs}

The merger of a ONe WD with a CO WD binary companion will result in a thick disk produced by the tidal disruption of the CO WD secondary about the more massive ONe primary. The temperatures and the densities within the disk may give rise to conditions susceptible to the initiation of carbon detonation.  Unlike the carbon-rich disk, however, the ONe core is not as easily detonated, and may survive the carbon detonation in the disk. We now quantitatively estimate the physical properties of this thick disk from fundamental physical considerations.

The temperature of the disk is approximately virial, $T (r) = f {GM_{1}m_{p}}/{k_{B} r}$, where $M_1$ is the mass of the primary, $r$ is the cylindrical radial distance in the disk, and $f = 0.2$ is a factor determined from SPH simulations of merging WDs \citep {Zhu_2013}. We assume the disk extends from the tidal disruption radius $R_{\rm tid} = q^{-1/3} R_2$ outwards. Here $q = M_2 / M_1$ is the mass ratio of the secondary to the primary, and $R_2$ is the radius of the secondary.

Simulations further demonstrate that the tidally-disrupted disk satisfies a steep surface density power-law scaling with cylindrical radius $r$,  $\Sigma (r) \propto r^{-\eta}$, and that the primary remains largely intact apart from a narrow range of mass ratios very close to unity \citep {Zhu_2013, kashyapetal15}.  For a typical surface density scaling exponent $\eta = 3$, the normalized surface density of a disk extending from the tidal disruption radius outward is $\Sigma (r) = (M_2 / 2 \pi R_{\rm tid}^2) (R_{\rm tid} / r)^3$ \citep {kashyapetal15}. The sound speed, as determined from the virial temperature, is $c_s (r) = (f \gamma G M_1 / r)^{1/2}$. In turn, the sound speed, combined with with the nearly-Keplerian angular velocity $\Omega (r) \simeq (G M_1 / r^3)^{1/2}$ in the inner disk yields the characteristic disk scale height   $h (r) = c_s (r) / \Omega (r) \propto r$, implying the disk is flared.

The temperature at the tidal disruption radius in the disk is $T (R_{\rm tid} ) =  f q^{1/3} GM_1 m_p /{k_B R_2}$, and the midplane density at the same radius is $\rho (R_{\rm tid}) = \Sigma (R_{\rm Tid}) / h (R_{\rm tid}) = (1 / f \gamma)^{1/2} (M_2 / 2 \pi R_{\rm tid}^3)$. For a $1.2$ $M_{\odot}$ ONe WD  primary with radius $\sim 5 \times 10^{-3} R_{\odot}$ \citep{tremblayetal17}, merging with a massive CO secondary ($q \sim 1$), the virial temperature at the tidal disruption radius is estimated to be $2 \times 10^{9}$ K, and the midplane density there $\sim 10^7$ g cm$^{-3}$. More massive ONe primaries lead to higher disk temperatures and densities still. Thus, the merger of a massive CO WD secondary with a ONe primary naturally leads to thermodynamic conditions in the inner tidally-disrupted disk favorable to carbon detonation \citep {Seitenzahl_2009, jordanetal12a, fisheretal18}.  In contrast, oxygen detonation under similar thermodynamic conditions leads to critical detonation length scales up to six orders of magnitude greater than carbon \citep {shenbildsten14}, potentially exceeding the ONe WD radius itself, and making the detonation of the ONe core a much more difficult prospect.

These analytic estimates pose additional questions: if a detonation is ignited in the carbon-rich disk, what is the fate of the ONe core? What are the nucleosynthetic yields of the burning? Are radioactive isotopes including $^{56}$Ni  produced, and if so, what are the observational characteristics of the transient? Motivated by these questions, in this paper, we report on a 3D simulation of a 1.2 $M_{\odot}$ ONe and 1.1 $M_{\odot}$ CO  binary WD merger. This paper is organized as follows: in section \ref {methodology}, we describe the methodology employed to sample the model parameter space of WD progenitors. In section \ref {results}, we present the results of our simulations, including nucleosynthetic yields and estimated observable properties. In sections \ref {discussion}  we discuss our findings and conclude.

\section {Methodology}
\label {methodology}

We consider a binary merger model consisting of a $1.2 M_{\odot}$ ONe primary WD with 80 percent O and 20 percent Ne, and a $1.1 M_{\odot}$ CO secondary WD with 40 percent carbon and 60 percent oxygen.  Our binary WD simulations utilize a two-stage simulation workflow. The first stage involves smoothed particle hydrodynamics (SPH) simulations, and follows the early evolution through the initial phase of the tidal disruption of the secondary. Like other similar white dwarf mergers, there is a central white dwarf and a tidally-disrupted accretion disk from the secondary. The SPH simulations are then remapped {\tt FLASH} adaptive mesh refinement code. These two code frameworks naturally complement one another. In particular, the SPH simulations, which lack a grid structure, are well-positioned to track the early dynamical phases of evolution. The Eulerian mesh simulations complement the angular-momentum conserving properties of the SPH evolution with the ability to accurately capture  hydrodynamic instabilities at high resolution \citep{2017ApJ...840...16K}.

These SPH simulations employ the code of \citet{Loren_Aguilar_2010} with $4 \times 10^5$ SPH particles. In the SPH simulations, nuclear reactions are modeled using 14  $\alpha$-capture nuclei from He to Zn, coupled with a quasi-equilibrium-reduced $\alpha-$network for temperatures higher than $3 \times 10^9$ K \citep{Hix_1998A}. The {\tt REACLIB} database \citep{Cyburt_2010} is employed for nuclear reaction rates and \cite{Itoh_1996} is used for neutrino loss rates, although the neutrino energetic losses are negligible over the timescale of the simulation. Additionally, because the simulation presented here is evolved for a dynamical time, and accretion evolves over a significantly longer timescale \citep {shenetal12}, the effect of the magnetorotational instability is not significant, and MHD is not included. However, absent the effect of the magnetic field, which effectively brings the primary into a state of uniform rotation \citep {jietal13} on the timescale of hours, here the primary remains differentially-rotating. Further, the hot magnetized corona seen in \citet {jietal13} is absent here. Both the SPH code and the {\tt FLASH}  code employ the same equation of state routine, with contributions from electrons, ions, and photons with an arbitrary degree of degeneracy and relativity \citep{Timmes_2000}.

We set up the initial SPH conditions to simulate the merger of the two WDs in a synchronized binary orbit. We further ensure accurate mass transfer as prescribed in \citet{2017ApJ...840...16K}.
In particular, we first set the merging WDs in a co-rotating frame at a  large distance, making sure the orbital shrinkage time is significantly larger than the dynamical timescale of the secondary WD, and accordingly that orbital separation is slowly reduced. The simulation of the merger starts as soon as the first SPH particle of the secondary WD fills its Roche lobe. 

The SPH data is then mapped onto {\tt FLASH} at 1.5 outer disk dynamical times \citep{2017ApJ...840...16K}. The upper left frame at $t=0$ s of Figure \ref {fig:figure_1} shows the density structure in the midplane at the time of remapping, $320$ s of evolution into the SPH model. We have explored the sensitivity of the outcome to the choice of the time of remapping, with a second remapped model at $340$ s; both results were very similar. {\tt FLASH} uses a 3D Eulerian grid with adaptive mesh refinement (AMR) for setting up hydrodynamic variables \citep{Fryxell_2000}. Gravity is computed using a multipole solver through $l = 60$ poles \citep{Couch_2013}. Within {\tt FLASH}, a 19 isotope $\alpha$-chain network is used to model nuclear burning \citep{Timmes_1999}. 

Our domain extends from $-2.8\times 10^{10}$ cm to $+2.8\times 10^{10}$~cm in each direction. We use a combination of mass and temperature criteria such that any {\tt FLASH} block containing one or more cells with mass exceeding $2\times 10^{-7} M_{\odot}$, and temperature exceeding $10^8$ K will be refined further, up to a maximum $6$ levels of resolution. On the finest level, the cell spacing corresponds to $136$ km.  We have also explored a variation on these criteria, using only the mass criterion. This resolution criterion variation resulted in a different grid distribution with equivalent finest resolution. The physical outcomes, including the nucleosynthetic yields, of the simulations were insensitive to this variation, however.

\section {Results}
\label {results}

\subsection {Hydrodynamic and Thermodynamic Time Evolution of the Binary WD Merger}

\begin{figure*}[h]
\begin{center}
\includegraphics[width=2\columnwidth]{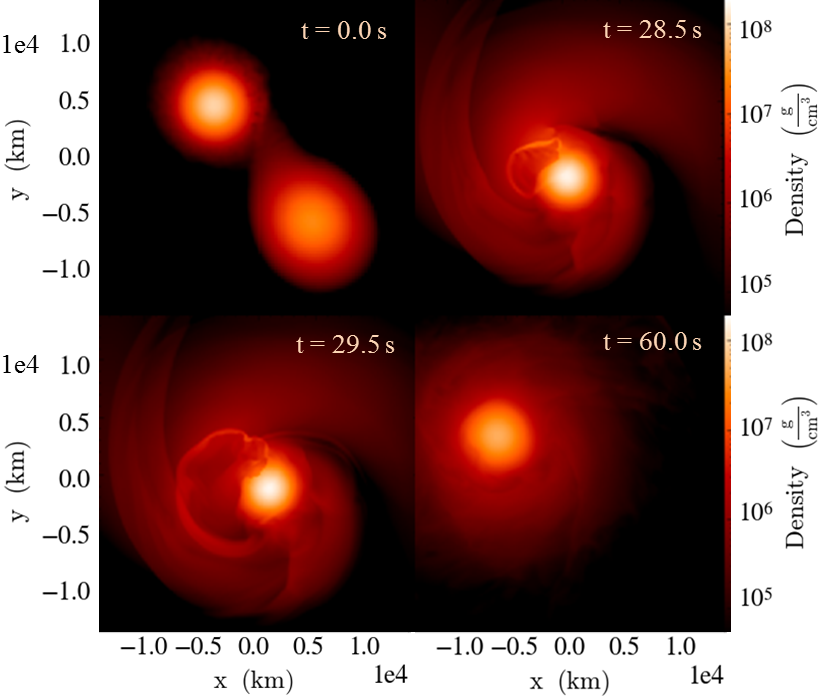}
\caption{Hydrodynamic time evolution of the binary WD system from the beginning of simulation to 60 s post merger. This figure represents the time evolution of the density in the mid-plane of the disk, with density shown on a logarithmic scale. The upper-left frame at $t$ = 0 s depicts the onset of unstable mass transfer from the secondary WD to the primary. The frame at top right taken $t$ = 28.5 s shows the development of a off-center hotspot corresponding to a density of $4 \times 10^{7}$ g cm$^{-3}$. The final frame at bottom right at $t$ = 60.0 s represents the remnant of the merger after a small amount of mass  ($\sim 0.08$ $M_{\odot}$) leaves the system as ejecta. All four frames are zoomed-in by a factor of 20 from the problem domain.}
\label {fig:figure_1}
\end{center}
\end{figure*}

\begin{figure*}[h]
\begin{center}
\includegraphics[width=2\columnwidth]{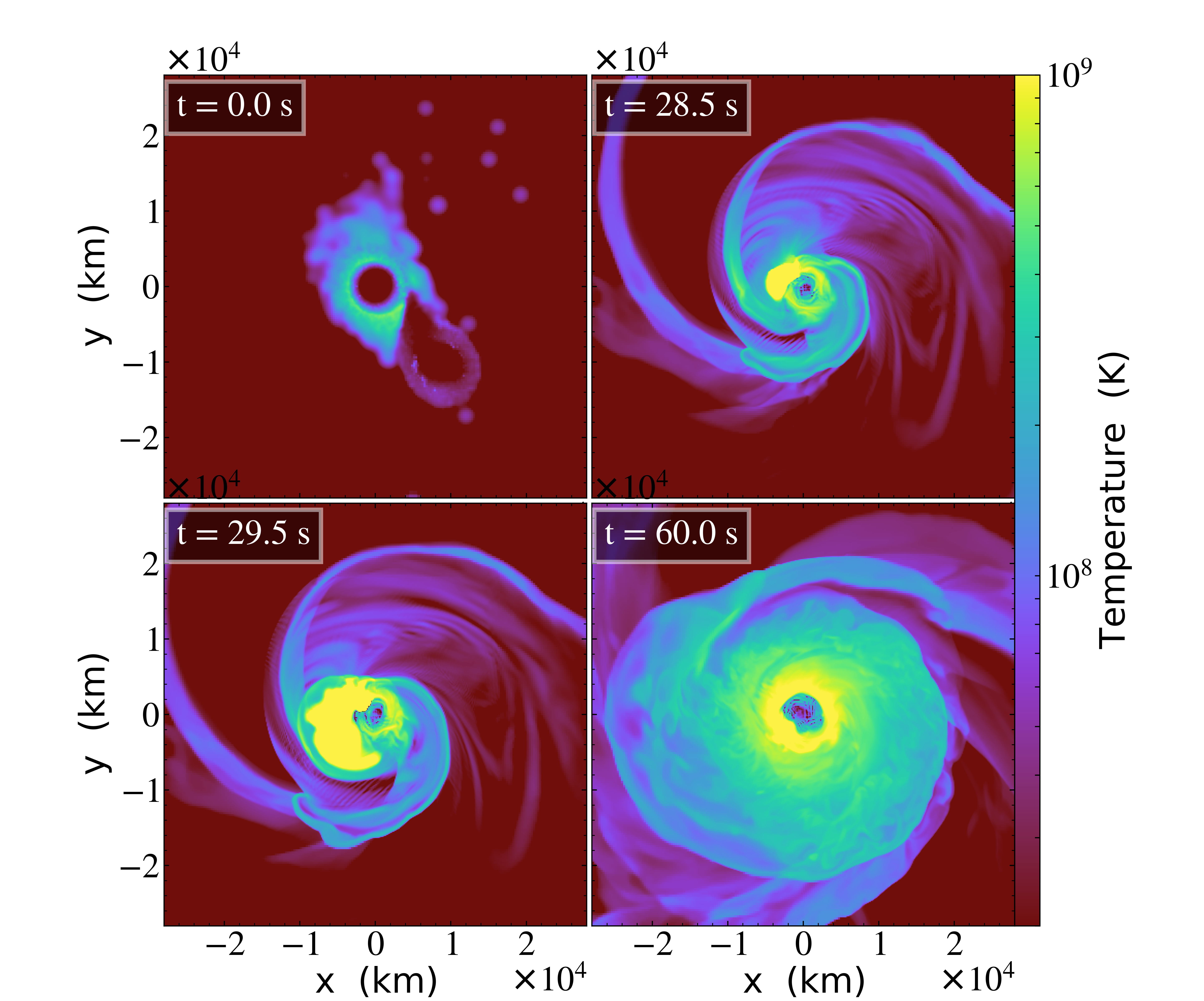}
\caption{Thermodynamic time evolution of the log temperature in the mid-plane of the disk, at the same times as shown in Figure \ref {fig:figure_1}. The core of both WDs are cold initially, with temperatures $T \simeq 10^{7}$ K. However, the CO secondary is rapidly tidally disrupted, forming a hot disk surrounding the primary ONe WD, and the onset of off-center carbon detonation peaked at $T$ $\sim$ $3.2\times 10^{9}$ K. The third frame at $t$ = 29.5 s illustrates the propagation of the detonation. All four of these frames are zoomed-in by a factor of 8 from the problem domain.}
\label {fig:figure_2}
\end{center}
\end{figure*}

Our {\tt FLASH} simulation begins with unstable mass transfer.  Figures \ref {fig:figure_1} and \ref{fig:figure_2} show the hydrodynamic and thermodynamic time-evolution of the binary, up to 60 s post-merger, depicting the log density and log temperature of the system in the midplane of the simulation domain. Note that the coordinate origin is located at the center-of-mass of the system, so that the primary WD is slightly off-centered with respect to the center-of-mass. The final inspiral of the binary WD merger begins as the secondary WD approaches the primary, initiating unstable mass transport, as depicted in the upper left frame at $t =0$ s of Figure \ref {fig:figure_1}. The secondary forms a hot, low-density accretion disk surrounding the primary.  An off-centered hotspot at the interface of the accretion disk and the primary, with a mixed composition of primary and secondary material ($X_{\rm O} = 0.62$, $X_{\rm C} = 0.22$, and $X_{\rm Ne} = 0.05$), is subsequently generated. The second frame of figure \ref {fig:figure_1}, at $t$ = 28.5 s, shows the development of this off-centered hotspot resulting from to the coalescence.  This coalescence leads to unstable carbon burning, and the subsequent initiation of carbon detonation at a temperature $T$ $\sim$ $3 \times 10^{9}$ K at a density of $\rho \sim 4 \times 10^{7}$ g cm $^{-3}$.  \\


\subsection{Accretion Disk Carbon Detonation Initiation}

Carbon detonation initiation arises on length scales substantially smaller than the typical resolution of three-dimensional numerical simulations \citep {seitenzahletal09}. For example, at a typical density and temperature of $10^7$ g cm$^{-3}$ and $2 \times 10^9$ K at which carbon burning enters into the distributed burning regime and a detonation is believed to be possible, the Zel'dovich gradient mechanism yields critical length scales of order 1 km. These critical length scales are highly sensitive to density and even more so to temperature, owing to the strongly nonlinear dependence of the carbon burning rate to these quantities. As a result, while carbon detonation initiation is unresolved in three-dimensional simulations, if the highest densities and temperatures achieved are sufficiently large, the carbon detonation critical lengths are sufficiently small that any disordered small-scale temperature distribution is likely to yield a physical detonation. Consequently, we require that the density exceeds  $10^7$ g cm$^{-3}$ and the  temperature exceeds $2 \times 10^9$ K  within a resolved hot spot in order to ensure a physical detonation. 
 
 We carefully examined the development of the hotspot which initiates the carbon detonation. We isolate the time of initiation by  the sharp increase in peak temperature at $t = 28.1$ s, signaling the onset of unstable carbon burning. The resolution of the hotspot was determined using a  volume-averaged spherical profile about the maximum temperature at the onset of unstable nuclear burning. The temperature profile reveals a marginally-resolved asymmetric hotspot spread over  $\sim$ 64 finest cells, and with a radius of $\sim$ $544$ km. 


While the critical detonation length scales for lower carbon fraction are significantly larger than for equal carbon-oxygen mixtures \citep {Seitenzahl_2009}, the density and temperatures reached in our case yield carbon detonation critical lengths that are sufficiently small that we deem the detonation physical. We consider the ramifications of the carbon detonation in the following.

\subsection{Failed Detonation of ONe WD}

The carbon detonation erupts at $t = 28.5$ s and rapidly burns through the carbon-rich disk. Crucially, only the innermost portion of the disk is at densities above $10^7$ g cm$^{-3}$, and is degenerate at the temperature of the detonation $\sim 10^9$ K \citep {timmeswoosley92}. Only material above densities of $\sim 10^7$ g cm$^{-3}$ synthesizes a significant amount of $^{56}$Ni \citep {nomotoetal84}. Consequently, the nucleosynthesis of the disk naturally leads to a very small yield of $^{56}$Ni. 


The third frame of Figure 3.1, taken at $t$ = 29.5 s post-merger, shows the propagation of the carbon detonation as it burns through the inner carbon-rich disk.  A small nuclear energy is released, ($E_{\rm nuc} = 3.2 \times 10^{50}$ erg), much less than the gravitational binding energy, and resulting in a failed detonation of the system. The ejecta propagates outward with a kinetic energy of $E_{\rm kin} = 2.2 \times 10^{49}$ erg. A amount of mass from the failed detonation($\sim$ $0.08$ $ M_{\odot}$), with sufficient energy to overcome gravitational potential is expelled from the system, leaving behind an intact ONe massive remnant bound core of $\sim$ $2.22$ $M_{\odot}$, as is illustrated in the fourth frame of Figure 3.1 at $t$ = 60 s post-merger. 
 
While the detonation propagates through the carbon-rich disk, the ONe primary remains largely intact and does not undergoes a nuclear detonation. This is because oxygen is far less susceptible to nuclear detonation than the carbon. \citet{shenbildsten14} showed that the critical radii for ONe detonations may be up to $10^{6}$ times larger than that found for CO detonations by \citet{Seitenzahl_2009}. In other words,  it is very challenging to initiate oxygen detonation in the ONe core. While there is some mixing of carbon near the surface of the ONe primary from the tidally-disrupted secondary WD, in this case it is insufficient to initiate oxygen detonation, and the result is the failed detonation of the ONe primary. 




\subsection{Post-Detonation Evolution, Fallback, and Ejected Mass} 
  The outburst is intrinsically asymmetric, and gives rise to an asymmetric ejecta distribution. A 3D volume rendering of the ejecta is shown in Figure \ref{fig:figure_4}. It is evident that the mass is ejected preferentially in the polar directions. As a result of the asymmetry of the model ejecta, the transient will present a viewing-angle-dependent light curve to the observer. 
 
 
 During the asymmetric outburst, the remnant is kicked at a velocity of $\sim$ $90$ km/s. The ejecta are launched in the opposite direction, maintaining momentum conservation. The mass-weighted root-mean-squared velocity of the unbound ejecta is calculated to be $\sim$ $4000$ km/s.
 
\begin{figure*}[]
\begin{center}
\includegraphics[width=2\columnwidth]{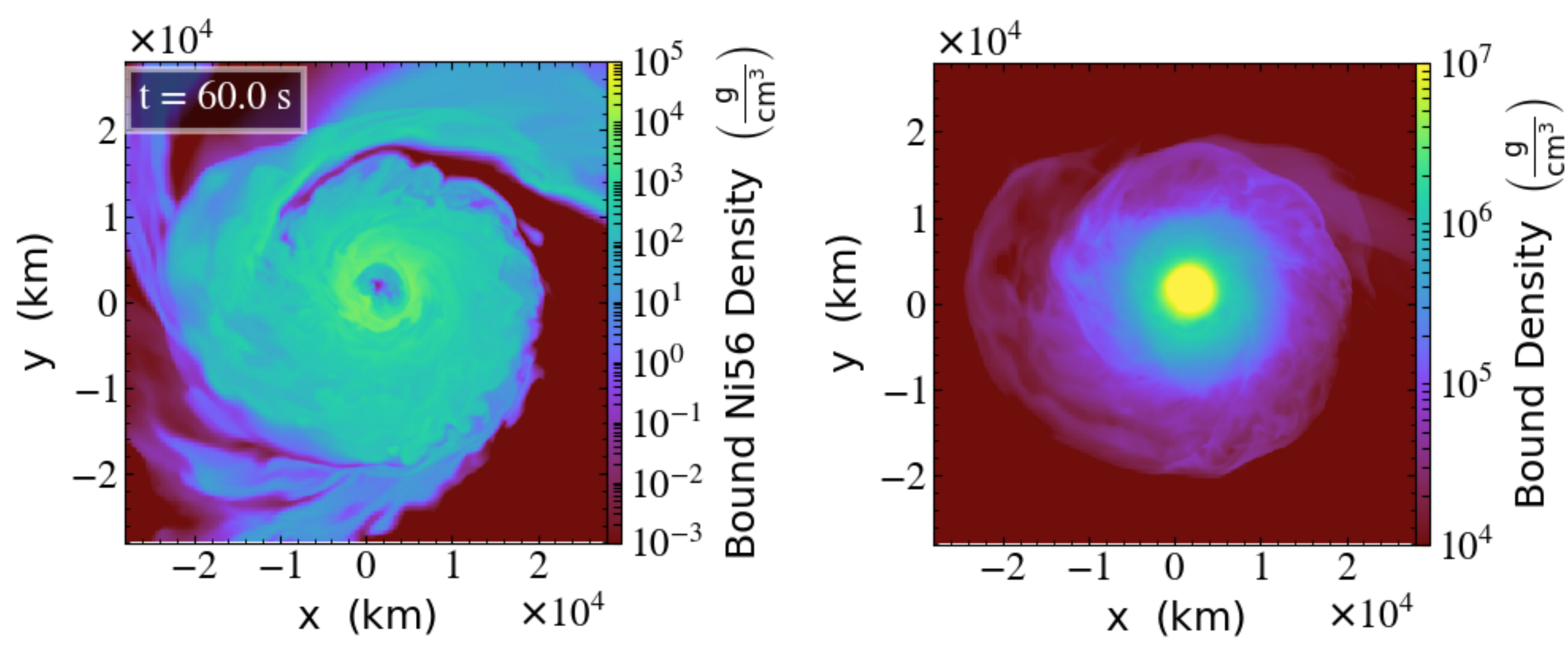}

\caption{Log of gravitationally-bound density and gravitationally-bound ${}^{56}$Ni density of the merged WD system in the mid-plane of the disk, at 60 s post-merger. Both frames are zoomed-in by a factor of 13 from the problem domain.}
\label {fig:figure_3}
\end{center}
\end{figure*}

\begin{figure*}[]
\begin{center}
\includegraphics[width=2\columnwidth]{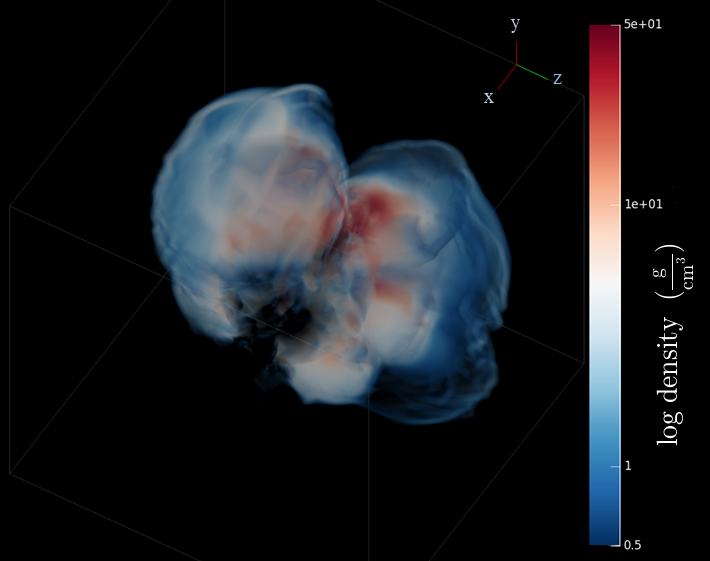}

\caption{A volume rendering of the ejecta density at $t$ = 60 s. The asymmetry of the ejecta in different directions is clearly visible, with the polar velocities exceeding those in the midplane.}
\label {fig:figure_4}
\end{center}
\end{figure*}

 
 However, not all of the burned material is unbound. The gravitationally-bound material surviving the outburst will fall back onto the remnant. Figure \ref{fig:figure_3} illustrates the bound density, and ${}^{56}$Ni density, respectively, at 60 s post merger, on a logarithmic scale in the midplane of the disk. The ${}^{56}$Ni distribution reveals that about half of the  ${}^{56}$Ni synthesized in the outburst falls back onto the disk and the remnant, and  will remain gravitationally bound to the system. While this simulation was terminated at $t = 60$ s, eventually nearly all of the  material in the disk, enriched by the fallback, will be accreted over a timescale of hours \citep {Ji_2013}.

We locally define the specific total energy density of the fluid as $E = (1/2) v^2 + e_{\rm int} + \Phi$. Assuming we may neglect additional sources of energy (such as radiatively-driven winds, possibly relevant at later times), we define the gas to be unbound when $E > 0$. Here $(1/2) v^2$ is the specific kinetic energy density of the gas, $e_{\rm int}$ is the specific internal energy, $\Phi$ is the gravitational potential. Gas with $E < 0$ is determined to the bound, and consists of the gravitationally-bound remnant including the bound accretion disk, as well as any fallback material which will not be ejected from the system. This gravitationally-boundedness criterion is used in the following to mask the gas into ejected and bound mass, respectively. 

As an outcome of the failed detonation, a very small amount of mass is burned ($\sim$ $2.6 \times 10^{-1}$ $M_{\odot}$), and a small amount of mass ($\sim 0.08$ $M_{\odot}$) is ejected from the system with a small kinetic energy ($\sim$ $10^{49}$ erg). The remaining nuclear ash, rich in IMEs and with a trace amount of ${}^{56}$Ni, falls back into the remnant. 

\begin{figure*}[]
\begin{center}
\includegraphics[width=1.5\columnwidth]{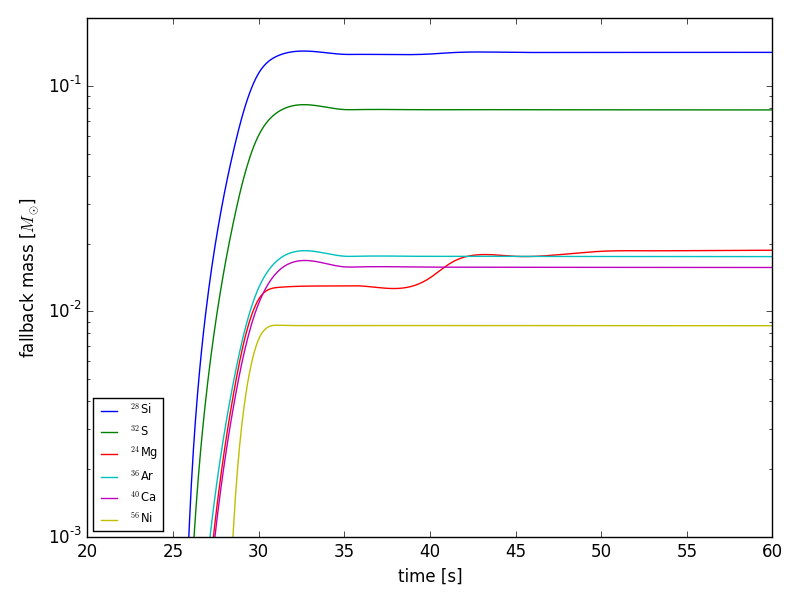}

\caption{Log of the fallback burned mass yields of the binary WD merger in solar masses versus time. Unburned $^{12}$C and $^{16}$O are not shown. The fallback is rich in intermediate-mass elements, in particular ${}^{28}$Si and ${}^{32}$S, with a trace amount of radioactive ${}^{56}$Ni. }
\label {fig:figure_5}
\end{center}
\end{figure*}

\begin{figure*}[]
\begin{center}
\includegraphics[width=1.5\columnwidth]{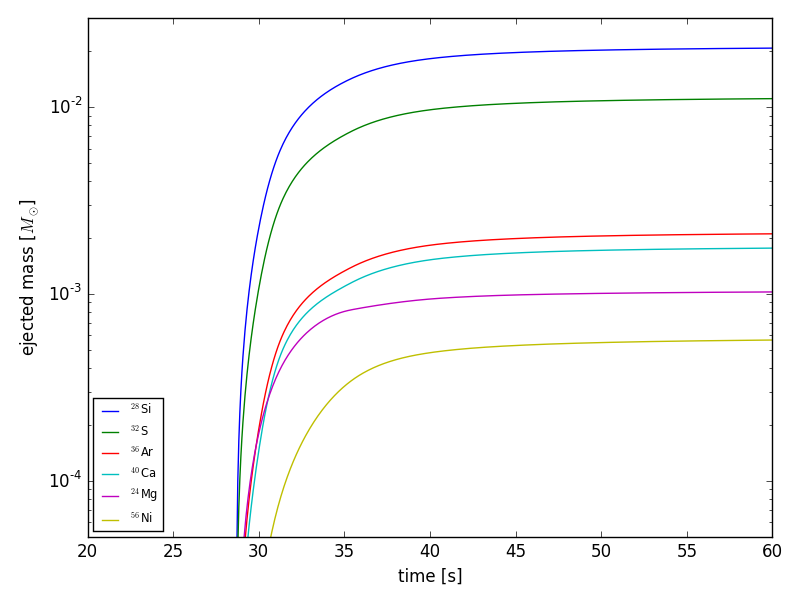}

\caption{Log of the ejected burned mass yields of the binary WD merger over the same times as shown in Figure 3.5. Unburned $^{12}$C and $^{16}$O are not shown. Similar to fallback materials, the ejecta are rich in intermediate-mass elements.}
\label {fig:figure_6}
\end{center}
\end{figure*}

Figures \ref{fig:figure_5}  and \ref{fig:figure_6} show the remnant and ejecta logarithmic mass abundances, respectively. The remnant are defined to consist of all gravitationally-bound material -- namely, both the surviving ONe core plus all the fallback mass bound to it. Within the ejecta, we find $\sim$ $8.2 \times 10^{-3} \/ M_{\odot}$(IGEs), including $\sim$ $5.66 \times 10^{-4}$ $M_{\odot}$ ${}^{56}$Ni. The most abundant species are ${}^{16}$O, ${}^{28}$Si, ${}^{12}$C, ${}^{32}$S, ${}^{36}$Ar and ${}^{20}$Ca. The isotopic mass yields in both the bound fallback and the ejecta are shown in Table 1. 
The ejecta are rich in $^{12}$C, $^{16}$O, $^{28}$Si,   and $^{32}$S, with significant amounts of $^{20}$Ne,  $^{24}$Mg, $^{36}$Ar, and $^{40}$Ca.

\begin{deluxetable}{c c c }
\tablecaption{Nucleosynthetic yields (in $M_{\odot}$) of isotopes in ejecta and fallback at $t=60$ s.\label{simtable} \vspace*{15pt} }
\vspace{10pt}
\bigskip
\tablehead{\colhead {Isotopes} & \colhead {Ejecta} & \colhead {Remnant} }
\startdata
${}^{12}$C & $1.15 \times 10^{-2}$  & $3.31 \times 10^{-1}$  \\
${}^{16}$O & $2.95 \times 10^{-2}$  & $1.43 $  \\ 
${}^{20}$Ne & $1.28 \times 10^{-3}$  & $2.17 \times 10^{-1}$ \\
${}^{24}$Mg & $1.02 \times 10^{-3}$  & $1.77 \times 10^{-2}$ \\ 
${}^{28}$Si & $2.06 \times 10^{-2}$  & $1.2 \times 10^{-1}$  \\
${}^{32}$S & $1.11 \times 10^{-2}$  & $6.71 \times 10^{-2}$ \\
${}^{36}$Ar & $2.09 \times 10^{-3}$  & $1.54 \times 10^{-2}$ \\
${}^{40}$Ca & $1.75 \times 10^{-3}$  & $1.39 \times 10^{-2}$ \\
${}^{44}$Ti & $9.40 \times 10^{-7}$  & $1.53 \times 10^{-5}$ \\
${}^{48}$Cr & $5.46 \times 10^{-6}$  & $5.28 \times 10^{-5}$ \\
${}^{52}$Fe & $3.87 \times 10^{-5}$  & $4.97 \times 10^{-4}$ \\
${}^{54}$Fe & $2.80 \times 10^{-7}$  & $4.72 \times 10^{-6}$ \\
${}^{56}$Ni & $5.66 \times 10^{-4}$  & $8.07 \times 10^{-3}$ \\
\enddata
\label {table:massfrac}
\end{deluxetable}

\subsection{Bolometric Light Curve}
The bolometric light curve of SNe Ia is primarily powered by the radioactive decay chain of ${}^{56}$Ni $\rightarrow$ ${}^{56}$Co $\rightarrow$ ${}^{56}$Fe \citep{1967CaJPh..45.2315T, 1969ApJ...157..623C}. Our simulation shows that a total of $8.64 \times 10^{-3}$ $M_{\odot}$ of ${}^{56}$Ni has been produced during the detonation, of which a very small amount, $5.66 \times 10^{-4}$ $M_{\odot}$, is ejected from the system. This ejected $^{56}$Ni will power a rapidly-fading optical transient.

\begin{figure*}[]
\begin{center}
\includegraphics[width=2\columnwidth]{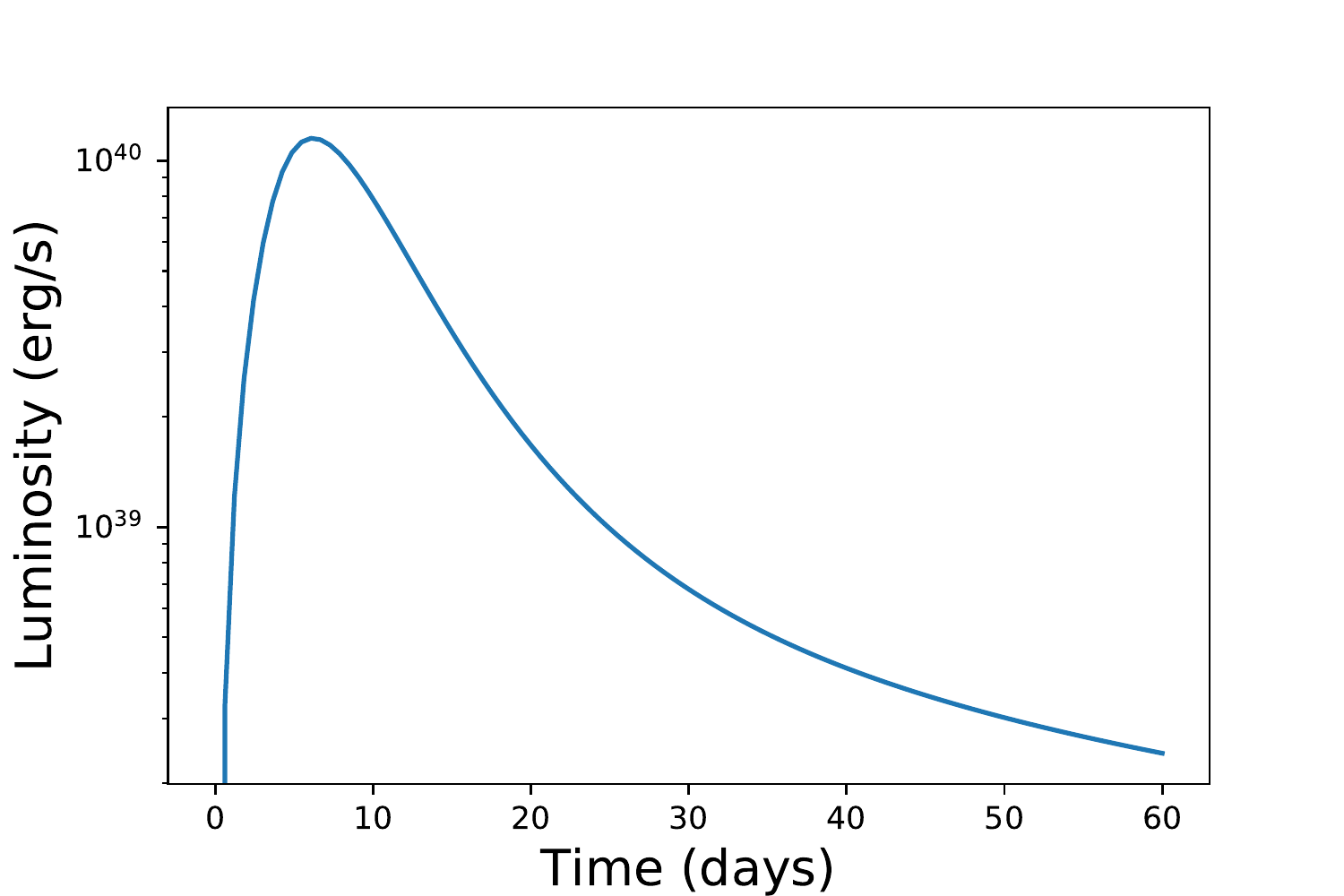}

\caption{Analytic bolometric light curve of the ONe and CO WD merger model using a simplified model using \cite{Dado2015}.}
\label {fig:figure_7}
\end{center}
\end{figure*}

The ejecta are asymmetric, and a full synthetic light curve and spectrum is desirable to ascertain the angle-dependence as seen by an observer. To ascertain the key observable signatures of the ONe-CO WD merger, we consider in this paper a simplified angle-averaged spherical approximation derived from an analytic model. In particular, in figure \ref{fig:figure_7} we show the bolometric light curve calculated from a simple analytic expression for bolometric luminosity \citep {Dado2015}. We use parameters relevant to our model, with ejecta mass $M_{\rm c} = 0.079$ $M_{\odot}$ and a ${}^{56}$Ni mass $8.6 \times 10^{-3}$ $M_{\odot}$. The fallback $^{56}$Ni will remain ionized and decay over much longer timescales \citep {shenschwab17}, and is therefore not included in the computation of the light curve. 
The bolometric luminosity peaks around day 6 post-explosion, which is comparatively earlier than the typical SNe Ia peak time, approximately 20 days post-explosion \citep{1999AJ....118.2675R}. 
We estimate our model peaks at bolometric magnitude $M \approx -11.4$.

\begin{figure*}[]
\begin{center}
\includegraphics[width=2\columnwidth]{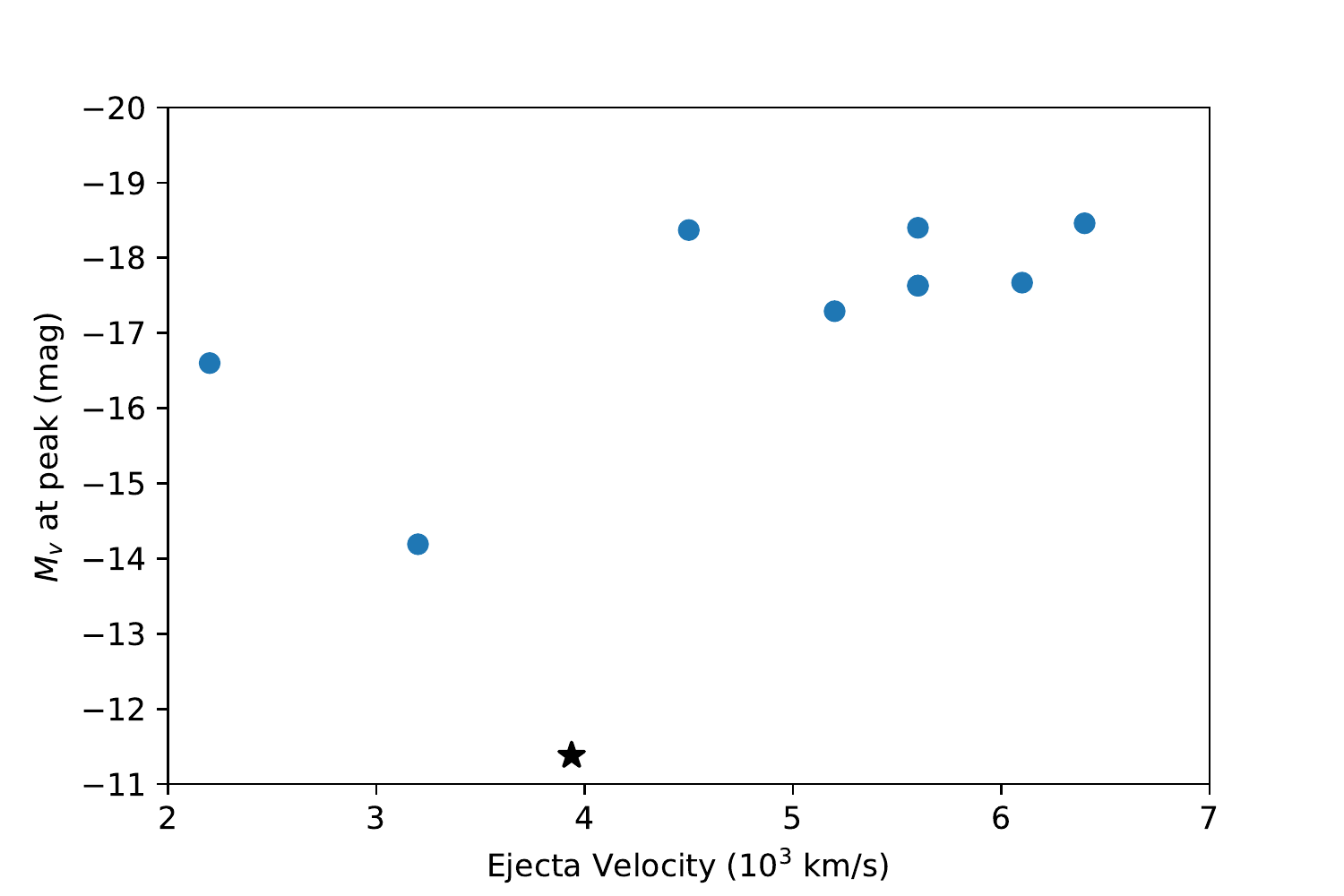}
\caption{Peak absolute magnitude vs. Si II velocity  for observed SNe Iax, shown in filled circles.
, along with estimated $M_{\rm peak}^{V}$ for SNe 2009J, 2009ku and 2012Z, presented with empty circles \citep{2013ApJ...767...57F}.  The velocity-magnitude relation of our ONe and CO WD merger is presented with a star. The very low ejecta velocity ($\sim$ $4000$ km/s) and faint luminosity ($M_V \sim$ $-11.3$) places the ONe and CO WD merger in an extremely faint and rapidly-declining region of the plot, fainter than the faintest SNe Iax yet observed. }
\label {fig:figure_8}
\end{center}
\end{figure*}


Figure \ref{fig:figure_8}  compares the velocity-magnitude of our theoretical model with some observed SNe Iax. Our peak magnitude and rise time are  fainter than the faintest and most rapidly-declining Type Iax observed to date, such as SN 2008ha and SN 2010ae, which have peak V band magnitudes $M_{\rm V} \sim -14$ mag at $t \sim 10$ days \citep{foleyetal09b, stritzingeretal14, jha17}.

\section {Discussion and Conclusions}
\label {discussion}

The energetics, nucleosynthetic yields, and predicted light curve of a ONe WD-CO WD merger clearly points to an unusual SNe Ia transient. The question then arises, which class of observed transients are in best agreement with the model predictions, if any?

Observations have revealed the existence of Ca-rich transients, which have strong Ca lines relative to O lines in their nebular spectra \citep {kasliwaletal12}. In contrast, our model ejecta are rich in O relative to Ca. Furthermore, the velocity of the ejecta is less than that observed in typical Ca-rich transients $\sim 6 \times 10^3 - 10^4$ km/s. Thus, this particular ONe and CO WD merger does not appear to match key observed properties of Ca-rich transients. However, more massive ONe primaries may yield more powerful explosions with higher Ca to O ratios, and would be worthwhile investigating as possible Ca-rich transient progenitors.

Another major class of unusual SNe Ia transients are the low-velocity SNe Iax. The leading proposed model for low-velocity SNe Iax invokes the failed deflagration of a near-Chandrasekhar mass CO WD \citep{jordanetal12b, kromeretal13}. In this model, the WD undergoes a vigorous deflagration phase, resulting in a significant nuclear energy release, and a great enough expansion of the WD that it fails to undergo a detonation. The nuclear ash of the deflagration is buoyantly expelled from the interior of the WD, which remains largely intact, and forms a remnant which survives the outburst. The total amount of burned mass ($\sim 10^{-2} M_{\odot}$) is small in comparison to normal SNe Ia. Some of the burned material is unbound to the  remnant, while the remaining portion, rich in $^{56}$Ni and other iron-peak products, falls back onto the remnant.  The ejecta is powered by a small amount of $^{56}$Ni, and would be visible as a rapidly-fading optical transient, rich in intermediate-mass elements including Ca, with low velocities compared to normal Ia -- all properties of observed low-velocity SNe Ia. Moreover, recent work suggests that the hot, ionized fall-back $^{56}$Ni will slowly decay, and power a longer-lived transient tail of the light curve \citep{shenschwab17}, consistent with observations which indicate that SNe Iax fail to enter the nebular phase as in normal SNe Ia \citep {jha17}.

While the failed deflagration model provides a  promising explanation for SNe Iax, the theory faces challenges in confronting observations. First, the failed deflagration model requires extraordinary fine tuning of the initial ignition conditions. It has long been known that Chandrasekhar-mass WDs naturally produce a large amount of  $^{56}$Ni in comparison to normal SNe Ia if promptly detonated, and therefore a vigorous deflagration is required to pre-expand the WD and produce a normal brightness event. Producing a failed deflagration from a Chandrasekhar-mass WD requires even greater deflagration than normal events, with $\sim 10 - 10^2$ ignition bubbles simultaneously ignited \citep {jordanetal12b, kromeretal13}. Furthermore, at least one of these bubbles must be ignited at distances very close ($<$  20 km at a central density of $\rho_c = 2 \times 10^9$ g/cm$^3$) to the WD center \citep {fisherjumper15}.  While the ignition conditions are still not precisely understood for the full range of near-Chandrasekhar mass WD central densities and compositions, the most detailed simulations conducted to date have demonstrated that the ignition is likely to occur at a single point, significantly offset from the WD center \citep{2011ApJ...740....8Z}. Consequently, current theory suggests the most likely outcome of such offset single-ignition single-degenerate SNe Ia will be a bright event \citep {fisherjumper15}. Failed events are possible in the single-degenerate channel, but are either less likely than bright events, or otherwise require substantial modifications to current theoretical models. 


Assuming these theoretical challenges can be overcome, the synthetic observables obtained in simulations of failed deflagration in near-$M_{\rm ch}$ CO WDs  do closely agree with the observational signatures of brighter SNe Iax like SN 2002cx and SN 2005hk \citep{kromeretal13}. However, the ${}^{56}$Ni yields of these models are larger than those inferred for the faintest SNe Iax, such as SN 2008ha and SN 2010ae. In their recent work, \citet{2015MNRAS.450.3045K} proposed a new mechanism of deflagration in a near-$M_{\rm ch}$ hybrid CONe WD, where the deflagration quenches in the carbon-rich core of the WD before traveling to the ONe mantle. This early-stage quenching gives rise to a significantly lower ${}^{56}$Ni yield ($3.4 \times 10^{-3}$ $M_{\odot}$). 

In contrast, our model indicates that the nucleosynthetic yield of a massive $1.1$ $M_{\odot}$ CO secondary with a  $1.2$ $M_{\odot}$ ONe primary in the double-degenerate channel naturally produces a failed detonation of the ONe core with a small ${}^{56}$Ni yield of $5.7 \times 10^{-4}$ $M_{\odot}$. The predicted outcome for this merger is a faint and rapid light curve fainter than the least luminous SNe Iax observed to date, such as SN 2008ha or SN 2010ae \citep {jha17}. The ONe and CO WD merger channel therefore naturally accommodates extremely faint SNe Ia. Consequently, the merger channel of ONe and CO WDs, along with predictions from other channels, such as the merger of neutron stars with WDs \citep {zenatietal18}, suggests that extremely rapid and faint transients may be detected in future surveys. Additionally, brighter events than the model computed here are also possible, since higher mass ONe primaries will naturally lead to higher virial temperatures and densities within the disk, and consequently greater ejecta masses and $^{56}$Ni yields. Further work exploring the diversity of the ONe and CO WD merger channel, and delving into the synthetic model spectra is warranted.

Recent work suggests that the distribution of weakly-bound material in WD mergers may produce fallback which may power optical emission on the timescale of  roughly a week \citep {ruedaetal18}. In contrast, the outburst here is sufficiently weak that the fallback occurs over a much shorter timescale, and is largely completed even by the end of the simulation presented here. Over a thermal timescale ($10^3 - 10^4$ yr), the evolution of the super-Chandrasekhar mass remnant surviving the outburst will likely result in a neutron star, similar to other merger remnants studied recently \citep {schwabetal16}. However,  it may also possible that the merger may result in an isolated Si WD if sufficient mass is driven off in winds. 


\citet {canalsetal18} have carried out binary population synthesis models and find that the ratio of ONe WD and CO WD mergers to the total SNe Ia rate, including both the double-degenerate and core-degenerate scenarios \citep {bearsoker18},  are in the range of 1\% - 10\%, depending on the assumed common envelope efficiency parameter. For comparison, the observed  rate of
SNe Iax events is $5$ to $30$ percent of the overall rate of SNe Ia \citep{2013ApJ...767...57F}. Consequently, from the standpoint of predicted rates alone, ONe and CO WD mergers  may account for a  subdominant fraction (3\% - 50\%) of all observed SNe Iax.

{\bf Acknowledgements.}  We dedicate this paper to the memory of E.G.-B., who tragically passed away just as the final paper was being completed. The work of P.L-A. and E.G.-B. was partially funded by the MINECO AYA2014-59084-P grant and by the AGAUR. RTF thanks the Institute for Theory and Computation at the Harvard-Smithsonian Center for Astrophysics for visiting support during which a portion of this work was undertaken. RTF acknowledges support from NASA 80NSSC18K1013. This work used the Extreme Science and Engineering Discovery Environment (XSEDE) Stampede 2 supercomputer at the University of Texas at Austin's Texas Advanced Computing Center  through allocation TG-AST100038, supported by National Science Foundation grant number ACI-1548562 \citep {townsetal14}.

\software {We utilize the adaptive mesh refinement code FLASH 4.0.1, developed by the DOE NNSA-ASC OASCR Flash Center at the University of Chicago. For nucleosynthetic post-processing we use TORCH, as described by \citet{timmes99}, and available from \url {http://cococubed.asu.edu/code_pages/net_torch.shtml}. For plotting and analysis, we have made use of yt \citep {Turk_2011}, \url {http://yt-project.org/}.}

\bibliography{converted_to_latex}



\end{document}